\documentclass[%
amsmath,amssymb,
aps,
twocolumn,
pra,
]{revtex4}
\usepackage{graphicx}
\usepackage{dcolumn}
\usepackage{bm}
\usepackage{xcolor}
\usepackage{booktabs}
\usepackage{xcolor}
\usepackage{framed}
\usepackage{multirow}
\usepackage{float}
\DeclareMathAlphabet \mathbfcal{OMS}{cmsy}{b}{n}

\begin{document}


\title{Topological Nanospaser}

\author{Rupesh Ghimire}
\email{rghimire1@student.gsu.edu}
\author{Jhih-Sheng Wu}%
 \email{b91202047@gmail.com}
\author{Vadym Apalkov}
\email{vapalkov@gsu.edu}
\author{Mark I. Stockman}%
 \email{mstockman@gsu.edu}
\affiliation{%
Center for Nano-Optics (CeNO) and Department of Physics and Astronomy, Georgia State University, Atlanta, Georgia 30303
}%

\date{\today}

\begin{abstract}
We propose a nanospaser made of an achiral plasmonic-metal nanodisk and a two-dimensional chiral gain medium -- a monolayer transition-metal dichalcogenide (TMDC). When one valley of the TMDC is selectively pumped (e.g., by a circular-polarized radiation), the spaser generates a mode carrying a topological charge (chirality) that matches that of the gain valley. The  chirally-mismatched, time-reversed mode has exactly the same frequency but the opposite topological charge; it is actively suppressed by the gain saturation (population clamping) and never generates leading to a strong topological protection for the generating matched mode. This topological spaser is promising for the use in nanooptics and nanospectroscopy in the near-field especially in applications to biomolecules that are typically chiral. Another potential application is a chiral nanolabel for biomedical applications emitting  in the far field an intense circularly-polarized coherent radiation.
\end{abstract}

\maketitle




\section{Introduction}
\label{Intro}

Spaser (surface plasmon amplificaiton by stimulated emission of radiation) was originally introduced in 2003 \cite{Bergman_Stockman:2003_PRL_spaser} as a nanoscopic phenomenon and device: a generator and amplifier of coherent nanolocalized optical fields. Since then, the science and technology of spasers experienced a rapid progress. Theoretical developments \cite{ Li_Li_Stockman_Bergman_PRB_71_115409_2005_Nanolens_Spaser, Stockman_JOPT_2010_Spaser_Nanoamplifier, Fedyanin_Opt_Lett_2012_Elecrically_Pumped_Spaser, Berman_et_al_OL_2013_Magneto_Optical_Spaser, Zheludev_et_al_Nat_Phot_2008_Lasing_Spaser} were followed by the first experimental observations of the spaser \cite{Noginov_et_al_Nature_2009_Spaser_Observation, Oulton_Sorger_Zentgraf_Ma_Gladden_Dai_Bartal_Zhang_Nature_2009_Nanolaser} and then by an avalanch of new developments, designs, and applications. Currently there are spasers whose generation spans the entire optical spectrum, from the near-infrared to the near-ultraviolet \cite{Zhang_et_al_Nature_Materials_2010_Spaser,  Long_et_al_Opt_Expr_2011_Spaser_1.5micron_InGaAs,  Hill_et_al_Opt_Expr_2011_DFB_SPP_Spaser, van_Exter_et_al_PRL_2013_Holy_Array_Spasing, Lu-2014-All-Color_Plasmonic, Xiong_et_al_ncomms5953_2014_Room_Temperature_Ultraviolet_Spaser, Lin_et_al_srep19887_2016_Single_Crystalline_Al_ZnO_UV_Spasers, Gwo_et_al_acsphotonics_7b00184_2017_Low_Threshold_Spasers, Song_et_al_acsphotonics_b01018_2017_Perovskite_Grating_Spaser}.

Several types of spasers, which are synonymously called also nanolasers, have so far been well developed. Historically, the first is a nanoshell spaser \cite{Noginov_et_al_Nature_2009_Spaser_Observation} that contains a metal nanosphere as the plasmonic core that is surrounded by a dielectric shell containing gain material, typically dye molecules \cite{Noginov_et_al_Nature_2009_Spaser_Observation, Galanzha_Nat_Comm_Spaser_biological_probe_2017}. Such spasers are smallest coherent generators produced so far, with sizes in the range of tens nanometers. Almost simultaneously, another type of nanolasers was demonstrated \cite{Oulton_Sorger_Zentgraf_Ma_Gladden_Dai_Bartal_Zhang_Nature_2009_Nanolaser} that was built from a semiconductor gain nanorod situated over a surface of a plasmonic metal. It has a micrometer-scale size along the nanorod.  Its modes are surface plasmon polaritons (SPPs) with nanometer-scale transverse size. Given that the spasers of this type are relatively efficient sources of far-field light, they are traditionally called nanolasers though an appropriate name would be polaritonic spasers. Later, this type of nanolasers (polaritonic spasers) was widely developed and perfected \cite{Zhang_et_al_Nature_Materials_2010_Spaser, Zhang_et_al_Nano_Lett_2012_Muliticolor_Spaser, Zhang_et_al_Nat_Nano_2014_Spaser_Explosives_Detection, Xiong_et_al_ncomms5953_2014_Room_Temperature_Ultraviolet_Spaser, Ma_acsphotonics.7b00438_2017_High_Stability_Spasers_for_Sensing, Wu_et_al-2018-Advanced_Optical_Materials_2018}. There are also spasers that are similar in design to the polaritonic nanolasers but are true nanospasers whose dimensions are all on the nanoscale. Such a spaser consists of a monocrystal nanorod of a semiconductor gain material deposited atop of a monocrystal nanofilm of a plasmonic metal \cite{Gwo_et_al_Science_2012_Spaser}. These spasers possess very low thresholds and are tunable in all visible spectrum by changing the gain semiconductor composition while the geometry remains fixed \cite{Lu-2014-All-Color_Plasmonic, Gwo_et_al_acsphotonics.7b00184_2017_Low_Threshold_Spasers, Gwo_Shih_RPP_2016_Semiconductor_Plasmonic_Nanolasers}. There are also other types of demonstrated spasers. Among them we mention semiconductor-metal nanolasers \cite{Ning-AP2019-Review_Nanolasers} and polaritonic spasers with plasmonic cavities and quantum dot gain media \cite{Kress-2017-Sci_Adv_2017_Quantum_Dot_Polaritonic_Spaser}.

A fundamentally different type of quantum generators is the lasing spaser  \cite{Zheludev_et_al_Nat_Phot_2008_Lasing_Spaser, Plum_Fedotov_Kuo_Tsai_Zheludev_Opt_Expr_2009_Toward_Lasing_Spaser, Zheludev_et_al_srep01237_2013_Torroidal_Lasing_Spaser}. A lasing spaser is 
a periodic array of individual spasers that interact in the near field and form a coherent collective mode. Such lasing spasers have been built of plasmonic crystals that incorporate gain media.  One type of the lasing spasers is a periodic array of holes in a plasmonic metal film deposited on a semiconductor gain medium \cite{van_Exter_et_al_PRL_2013_Holy_Array_Spasing}; another type is a periodic array of metal nanoparticles surrounded by a dye molecules solution \cite{Odom_et_al_Nature_Nano_2013_Spasing_in_Strongly_Coupled_Nanoprticle_Array}. We have recently proposed a topological lasing spaser that is built of a honeycomb plasmonic crystal of silver nanoshells containing a gain medium inside \cite{wu2019topological}. The generating modes of such a spaser are chiral surface plsmons (SPs) with topological charges of $m=\pm1$, which topologically protects them against mixing. Only one of the $m=\pm1$ topologically-charged modes can generate at a time selected by a spontaneous breaking of the time-reversal symmetry.

\begin{figure}
\begin{center}
\includegraphics[width=.95\columnwidth]{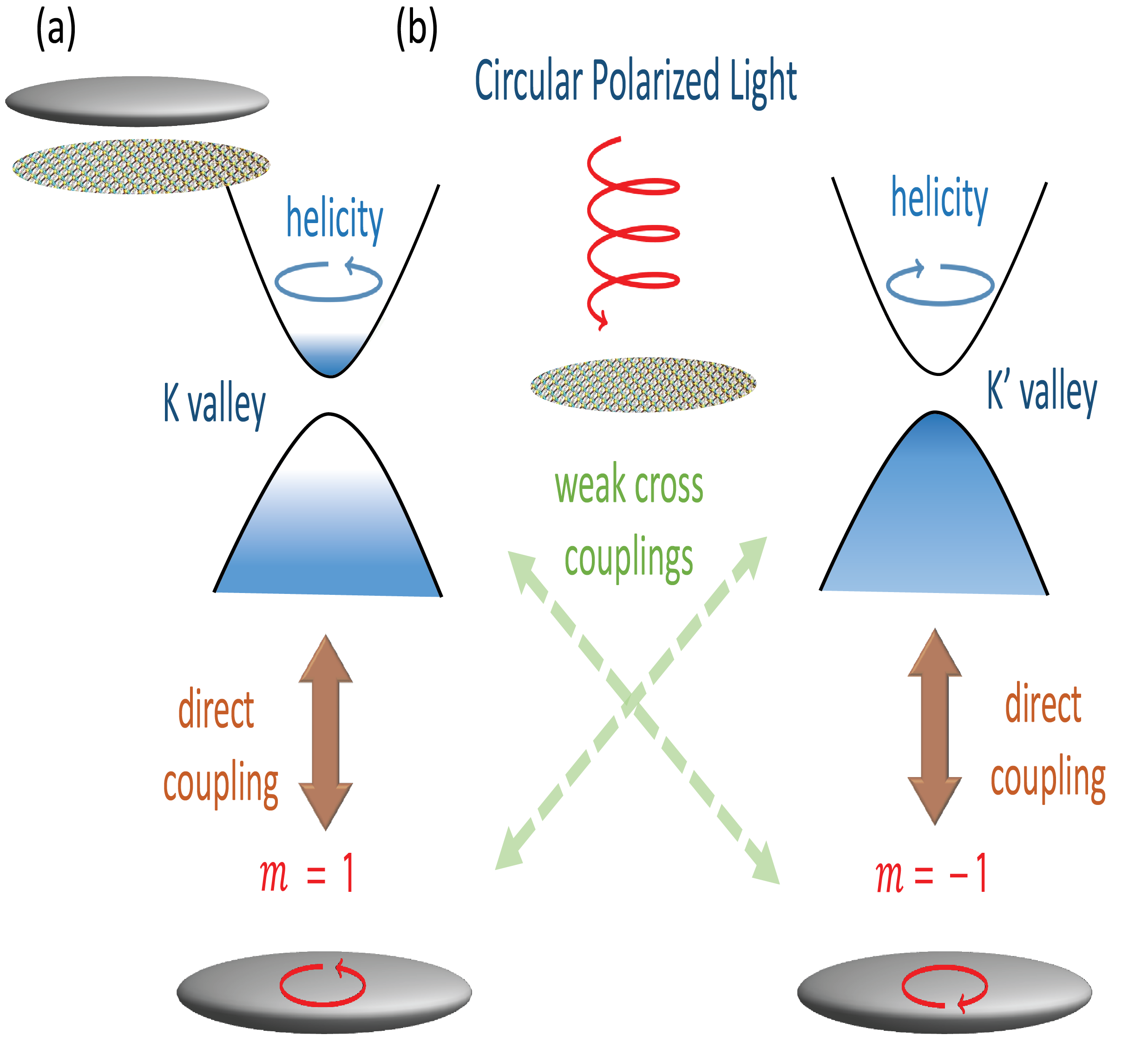}
\caption{Schematic of the spaser geometry and operation.
(a) Geometry of the spaser: Silver nanospheroid on TMDC. (b) Schematic of spaser operation. Pumping with a circular-polarized light excites the valley whose chirality is matched to the light helicity. The stimulated CB$\to$CB transitions at the corresponding $K$ or $K^\prime$ point excite SPs matched by chirality to that of the valley; the other, mismatched valley couples only weakly to these SPs.
}\label{fig:tmdc_spaser}
\end{center}
\end{figure}

The spasers are not only of a significant fundamental interest but also are promising for applications based on their nanoscale-size modes and high local fields. Among such demonstrated applications are those to sensing of minute amounts of chemical and biological agents in the environment \cite{Zhang_et_al_Nat_Nano_2014_Spaser_Explosives_Detection, Ma_acsphotonics.7b00438_2017_High_Stability_Spasers_for_Sensing, Ma_et_al_nanoph-2016-0006_Nanophotonics_2017_Spaser_Sensing}. Another class of the  demonstrated applications of the spasers is that in cancer theranostics (therapeutics and diagnostics) \cite{Galanzha_Nat_Comm_Spaser_biological_probe_2017}. An important perspective application of spasers is on-chip communications in optoelectronic information processing \cite{Stockman_2018_Patent_Processor_Communications}.

It is of a great interest to explore intersections of the spaser technology and topological physics. In our recently proposed topological lasing spaser \cite{wu2019topological}, the topologically-charged eigenmodes stem from the Berry curvature \cite{Berry_Phase_Proc_Royal_Soc_1984,  Xiao_Niu_RevModPhys.82_2010_Berry_Phase_in_Electronic_Properties} of the plasmonic Bloch bands of a  honeycomb plasmonic crystal of silver nanoshells. In contrast, the gain medium inside these nanoshells is completely achiral. This topological lasing spaser is predicted to generate a pair of mutually time-reversed eigenmodes carrying topological charges of $\pm1$, which strongly compete with each other, so only one of them can be generated at a time.

In this Article we propose a topological nanospaser that also generates a pair of mutually time-reversed chiral SP eigenmodes with topological charges of $\pm1$, whose fields are rotating in time in the opposite directions. In a contrast to Ref.\ \onlinecite{wu2019topological}, this proposed spaser is truly nanoscopic, with a radius $\sim10$ nm. The topological charges (chiralities) of its eigenmodes originate from the Berry curvature of the gain-medium Bloch bands. This gain medium is a two-dimensional honeycomb nanocrystal  of a transition metal dichalcogenide (TMDC)  \cite{Heinz_et_al_Nat_Phys_2015_Biexcitons_in_WSe2, Novoselov_et_al_Science_2016_2D_Materials_and_Heterostructures, Basov_et_al_aag1992.full_2016_Polaritons_in_2D}. The plasmonic subsystem is an achiral nanodisk of a plasmonic metal. Note that previously the TMDCs have been used as the gain media of microlasers where the cavities were formed by microdisk resonators \cite{Salehzadeh_et_al_ACSPublications_Optically_Pumped_2015_MoS2, Zhang_et_al_Nat_Phot_2015_Monolayer_Excitonic_Laser} or a photonic crystal microcavity \cite{Wu_et_al_Nature_2015_WSe2_PC_Nanocavity_Laser}. None of these lasers generated a chiral, topologically charged mode.

\section{Spaser Structure and Main Equations}
\label{Main_Eqs}

The geometry and the fundamentals of functioning of the proposed topological nanospaser is illustrated in Fig.\ \ref{fig:tmdc_spaser}. 
This spaser consists of a thin silver nano-spheroid placed atop of the two-dimensional (2d) gain medium (a nanodisk of a monolayer TMDC) -- see Fig.\ \ref{fig:tmdc_spaser}(a). 
As panel (b) illustrates, the gain medium is pumped with circularly-polarized light, which is known to selectively populate one of the $K$ or $K^\prime$ valleys depending on its helicity \cite{Cao_et_al_Nat_Commun_2012_Valley_Circular_Dichroism_MoS2, Heinz_et_al_Nat_Phys_2017_Optical_Manipulation_of_Valley_Pseudospin}. Due to the axial symmetry, the plasmonic eigenmodes, $\phi(\mathbf r)$, depend on the azimuthal angle, $\varphi$:  $\phi_m(\mathbf r)\propto \exp{(i m \varphi)}$, where $m=\mathrm{const}$ is the magnetic quantum number. Figure \ref{fig:tmdc_spaser}(b) illustrates that the conduction band (CB) to valence band (VB) transitions in the TMDC couple predominantly to the SPs whose chirality matches that of the valley: the transitions in $K$- or $K^\prime$-valley excite the $m=1$ or $m=-1$ SPs, respectively. 

The surface plasmon eigenmodes $\phi_{n}(\mathbf{r})$ are described by the quasistatic equation \cite{Stockman:2001_PRL_Localization}
\begin{align}
\nabla\Theta(\mathbf{r})\nabla \phi_{n} (\mathbf{r}) = s_{n}\nabla^2\phi_{n} (\mathbf{r}) ,
\end{align}
where $n$ is a set of the quantum numbers defining the eigenmode, $s_n$ is the corresponding eigenvalue, which is a real number between 0 and 1, and $\Theta(\mathbf{r})$ is the characteristic function, which is equal to 1 inside the metal and 0 outside. We assume that the metal nanoparticle is a spheroid whose eigenmodes can be found in oblate spheroidal coordinates \cite{Willatzen_Voon_2011_Book_Boundary_Problems} -- see Supplemental Materials (SM). They are characterized by two integer spheroidal quantum numbers: multipolarity $l=1,2,\dots$ and azimuthal or magnetic quantum number $m=0,\pm1,\dots$. We will consider a dipolar mode, $l=1$ where $m=0,\pm 1$. Note that the dipole transitions in the TMDC at the $K$-, $K^\prime$-points are chiral, and they couple only to the modes with $m=\pm1$. 

The Hamiltonian of the SPs is 
\begin{align}
H_{SP}=\hbar\omega_\mathrm{sp}\sum_{m=\pm1}  \hat{a}_m^{\dagger}\hat{a}_m^{},
\end{align}
where $\omega_\mathrm{sp}$ is the SP frequency, and $\hat{a}_m^{\dagger}$ and $\hat{a}_m^{}$ are the SP  creation and annihilation operators (we indicate only the magnetic quantum number $m$). The electric field operator  is \cite{Bergman_Stockman:2003_PRL_spaser, Stockman_JOPT_2010_Spaser_Nanoamplifier}
\begin{align}
\mathbf{F}_m(\mathbf{r},t)&= - A_\mathrm{sp} \nabla \phi_m(\mathbf{r})(\hat{a}_m e^{-i\omega_\mathrm{sp} t} +\hat{ a}_m^{\dagger} e^{i\omega_\mathrm{sp} t}),\label{Fm}\\
A_\mathrm{sp}& =\sqrt{\frac{4\pi\hbar s_\mathrm{sp}}{\epsilon_d s_\mathrm{sp}^\prime}}, \label{Asp}
\end{align}
where $s_\mathrm{sp}^\prime = \mathrm{Re}[{d s(\omega)}/{d\omega}|_{\omega=\omega_\mathrm{sp}}]$.
The monolayer TMDC is coupled to the field of the SPs via the dipole interaction. We choose the proper thickness of the silver spheroid so that 
the SP energy $\hbar\omega_\mathrm{sp}$  is equal to the band gap of the TMDC gain medium. The Hamiltonian of the TMDC near the $K$ or $K^\prime$ point can be written as
\begin{align}
H_{\mathbfcal{K}} =\int d^2\mathbf{q}\sum_{\alpha=\mathrm{v,c}} E_{\alpha}(\mathbfcal{K}+\mathbf{q})|\alpha,\mathbfcal{K}+\mathbf{q}\rangle\langle \alpha,\mathbfcal{K}+\mathbf{q}|,
\end{align}
where $\mathbfcal{K}= K$ or $ K^\prime$, and $\mathrm{v}$ and $\mathrm c$ stand for the valence band and the conduction band, correspondingly.
We expand the Hamiltonian around the $K$ and $K^\prime$ points as
\begin{align}
H_{\mathbfcal{K}} \simeq  {\nu}_{\mathbfcal{K}}\sum_{\alpha =c,~v} E_{\alpha}(\mathbfcal{K})
|\alpha,\mathbfcal{K}\rangle\langle \alpha,\mathbfcal{K}|,
\end{align}
where  $\nu_\mathbfcal{K}$ is the density of electronic states in the $\mathbfcal{K}$ valley, which we adopt from experimental data {\cite {Salehzadeh_et_al_ACSPublications_Optically_Pumped_2015_MoS2, Wu_et_al_Nature_2015_Monalayer}:
$  \nu_\mathbf{K} = \nu_\mathbf{K^\prime} =7.0\times10^{12}$~cm$^{-2} $.

The field of the SPs in nanoparticles  is highly nonuniform in space, which gives rise to a spatial non-uniformity of the electron population of the TMDC monolayer. To treat this, we employ a semiclassical approach where  the state $|\alpha,\mathbfcal{K},\mathbf{r}\rangle$  represents a electron in the $\mathbfcal{K}$ valley at position $\mathbf{r}$.  The corresponding  Hamiltonian in the semiclassical approximation can be written as 
\begin{align}
H_{\mathbfcal{K}}
 = \nu_{\mathbfcal{K}}\sum_{\alpha =c,~v}E_{\alpha}(\mathbfcal{K}) \int d^2\mathbf{r} 
|\alpha,\mathbfcal{K},\mathbf{r}\rangle\langle \alpha,\mathbfcal{K},\mathbf{r}|
\end{align}
The interaction between the monolayer TMDC and the SPs is described by an interaction Hamiltonian
\begin{align}
H_{int}=- \sum_{\mathbfcal{K}=\mathbf K, \mathbf K^\prime}\nu_{\mathbfcal{K}}\int d^2\mathbf{r}\sum_{m=\pm1}\mathbf{F}_m(\mathbf{r})\hat{\mathbf{d}}_\mathbfcal K(\mathbf r)~,
\end{align}
where the dipole operator is given by 
\begin{equation}
\hat{\mathbf{d}}_{\mathbfcal{K}}(\mathbf r)= \mathbf{d}_{\mathbfcal{K}}e^{i\Delta_\mathrm gt}|c,\mathbfcal{K},\mathbf{r}\rangle\langle v,\mathbfcal{K},\mathbf{r}| + \mathrm{h.c.}~,
\end{equation}
 where $\hbar\Delta_\mathrm g$ is the band gap (at the $K$- or $K^\prime$-point).
 
 The  transition dipole element, $\mathbf{d}_{\mathbfcal{K}}$, is related to the non-Abelian (interband) Berry connection $\mathbfcal A^\mathrm{(cv)}(\mathbf k)$ as
\begin{eqnarray}
\mathbf{d}_{\mathbfcal{K}}& =&e\mathbfcal A^\mathrm{(cv)}(\mathbf k)~, 
\nonumber\\
\mathbfcal A^\mathrm{(cv)}(\mathbf k)&=& i\left.\left\langle u_\mathrm c(\mathbf{k}) \left|
\frac{\partial}{\partial \mathbf{k}} \right|u_\mathrm v(\mathbf{k})\right\rangle\right|_{\mathbf{k}=\mathbfcal{K}}~,
\end{eqnarray}
where $u_\mathrm \alpha(\mathbf{k})$ are the normalized lattice-periodic Bloch functions.

In this Article, we consider the dynamics of the system semiclassically: we treat the SP annihilation and creation operators as complex $c$-numbers, $\hat{a}_m=a_m$ and $\hat{a}_m^\dag=a_m^\ast$, and describe the electron dynamics quantum mechanically by  density matrix $\hat{\rho}_{\mathbfcal{K}}(\mathbf{r},t)$. Furthermore,  we assume that the SP field amplitude is not too large, $\tilde{\Omega}_{m,\mathbfcal{K}} \ll \Delta_\mathrm g$, where the Rabi frequency is defined by 
\begin{align}
\tilde{\Omega}_{m,\mathbfcal{K}}(\mathbf{r})&= -\frac{1}{\hbar} A_\mathrm{sp} \nabla \phi_m(\mathbf{r})\mathbf{d}_\mathbfcal{K}^\ast~.
\label{Rabi}
\end{align}
Then we can employ the rotating wave approximation (RWA) \cite{Bloch_Siegert_PhysRev.57.522_1940, Agarwal_PhysRevA.4.1778_1971} where the density matrix can be written as
\begin{align}
\hat{\rho}_{\mathbfcal{K}}(\mathbf{r},t) & =\left(
\begin{array}{cc}
\rho_{\mathbfcal{K}}^\mathrm{(c)}(\mathbf{r},t) & \rho_{\mathbfcal{K}}(\mathbf{r},t) e^{i\omega t}\\
\rho_{\mathbfcal{K}}^{*}(\mathbf{r},t)  e^{-i\omega t} & \rho_{\mathbfcal{K}}^\mathrm{(v)}(\mathbf{r},t)
\end{array}
\right).
\end{align}
 Following Ref.\ \onlinecite{Stockman_JOPT_2010_Spaser_Nanoamplifier}, 
the equations of motion of the SPs and the monolayer TMDC electron density matrix are  
\begin{align}
\dot{a}_m=[i&(\omega - \omega_\mathrm{sp}) - \gamma_\mathrm{sp}]{a}_m +
\nonumber\\
& i \nu_\mathbfcal{K}\int_S d^2{\mathbf{r}}\sum_{\mathbfcal{K}}\rho^{*}_{\mathbfcal{K}}(\mathbf{r})\tilde{\Omega}_{m,\mathbfcal{K}}^{*}(\mathbf{r})~,
\label{12}
\end{align}
\begin{align}
\dot{n}_{\mathbfcal{K}}(\mathbf{r})=&-4\sum_{m=1,-1} \mathrm{Im} \left[{\rho}_{\mathbfcal{K}}(\mathbf{r}) \tilde{\Omega}_{m,\mathbfcal{K}}(\mathbf{r})a_{m}^{}\right] +\nonumber\\
&g_\mathbfcal{K} \left[1-{n}_{\mathbfcal{K}}(\mathbf{r})\right]-
\gamma_{2\mathbfcal{K}}(\mathbf{r}) \left[1 +{n}_{\mathbfcal{K}}(\mathbf{r})\right],~\label{13}~
\end{align}
\begin{align}
\dot{\rho}_{\mathbfcal{K}}(\mathbf{r})=[-i(\omega-\Delta_\mathrm g)-&\Gamma_{12}] {\rho}_{\mathbfcal{K}}(\mathbf{r})+
\nonumber\\
&i n_{\mathbfcal{K}}(\mathbf{r})\sum_{m=1,{-1}} \tilde{\Omega}_{m,\mathbfcal{K}}^{*}a_{m}^{*}~,
\label{14}
\end{align} 
where $S$ is the entire area of the TMDC, $\omega_\mathrm{sp}$ is the SP frequency,  $\gamma_\mathrm{sp}$ is the SP relaxation rate, $\Gamma_{12}$ is the polarization relaxation rate for the spasing transition $2\to1$, $g_\mathbfcal K$ is the pumping rate in valley $\mathbfcal K$,  the population inversion, $n_{\mathbfcal{K}}$, is defined as
\begin{equation}
n_{\mathbfcal{K}} \equiv \rho_{\mathbfcal{K}}^\mathrm{(c)} - \rho_{\mathbfcal{K}}^\mathrm{(v)}~,
\end{equation}
and the spontaneous emission rate of the SPs is \cite{Stockman_JOPT_2010_Spaser_Nanoamplifier}
\begin{align}
\gamma_{2\mathbfcal{K}}(\mathbf{r}) = \frac{2(\gamma_\mathrm{sp}+\Gamma_{12})}{(\omega_\mathrm{sp}+\Delta_\mathrm g)^2+(\gamma_\mathrm{sp}+\Gamma_{12})^2}\sum_{m=1,-1}\left| \tilde{\Omega}_{m,\mathbfcal{K}}(\mathbf{r})\right|^2
\end{align}

\section{Results and Discussion}
\label{Results}

\subsection{Parameters of Spaser and Chiral Coupling to Gain Medium}
\label{Params}

We consider a spaser consisting of an oblate silver spheroid with semi-major axis $a=12$ nm placed atop of a circular TMDC flake of the same radius. We assume that the system is embedded into a dielectric matrix with permittivity $\epsilon_\mathrm d=2$. We choose the value of the semi-minor axis $c$ (the height of the silver spheroid) to fit $\omega_\mathrm{sp}$ to the $K$-point CV$\to$VB transition frequency in the TMDC, $\omega_\mathrm{sp}=\Delta_\mathrm g$. We employ the three-band tight-binding model for monolayers of group-VIB TMDCs of Ref.\ \onlinecite {Liu_et_al_PRB_2014_Three_Band_Model}. We also set $\hbar\Gamma_{12} = 10$ meV. 

From the tight-binding model, we calculate the band structure, including band gap $\Delta_\mathrm g$ and the transition dipole matrix element $\mathbf d$. Note that at the $K$- and $K^\prime$-points, the band gaps are the same, $\Delta_\mathrm g\left(\mathbf K\right)=\Delta_\mathrm g\left(\mathbf K^\prime\right)$, while the transition dipole matrix elements are complex conjugated, $\mathbf d_\mathbf K=\mathbf d_{\mathbf K^\prime}^\ast$, as protected by the $\mathcal T$-symmetry. The values used in the computations are listed in the SM. Here we give an example for MoS$_2$: $c=1.2$ nm; $\hbar\Delta_\mathrm g=1.66$ eV; 
$\mathbf d_\mathbf K=17.7 ~\mathbf e_+~\mathrm D$, and $\mathbf d_{\mathbf K^\prime}=17.7 ~\mathbf e_-~ \mathrm D$, where $\mathbf e_\pm=\left(\mathbf e_x\pm i \mathbf e_y\right)/\sqrt2$ are chiral unit vectors.

\begin{figure}
	\begin{center}
		\includegraphics[width=.95\columnwidth]{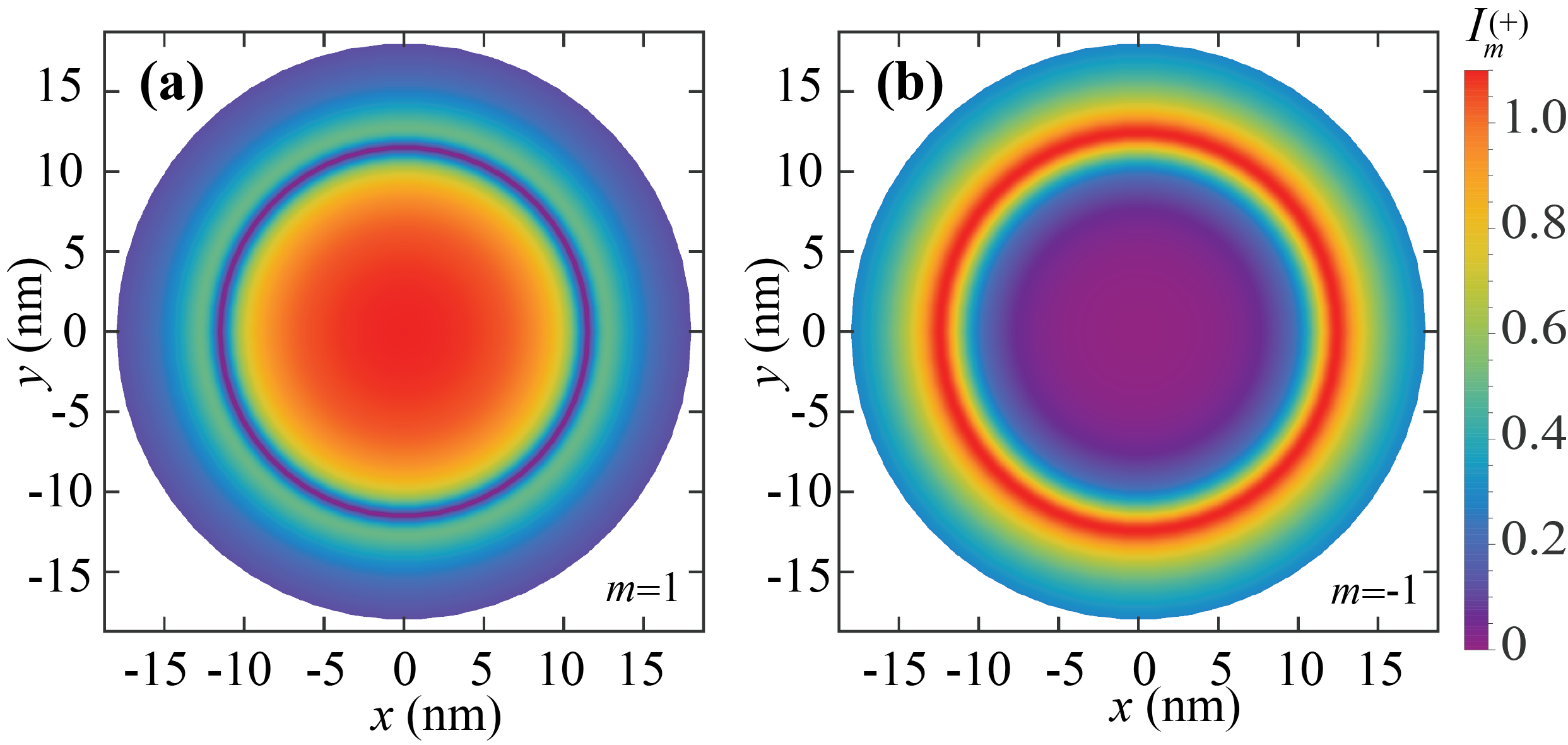}
		\caption{Coupling amplitude between SPs and TMDC dipole transitions, $I^{(+)}_m$, for $m=\pm1$ as indicated. The magnitude is color coded by the bar to the right. The radius of the metal spheroid is $a=12$ nm.}
		\label{SP_Coupling}
	\end{center}
\end{figure}

A fundamental question regarding any spaser is the existence of a finite spasing threshold. There are two modes with the opposite chiralities, $m=\pm1$, and identical frequencies, $\omega_\mathrm{sp}$, which are time-reversed with respect to each other, whose wave functions are $\nabla\phi\propto e^{\pm i\varphi}$. In the center of the TMDC patch, i.e., at $\mathbf r=0$, the point symmetry group of a metal nanospheroid on the TMDC is $C_{3v}$. It contains a $C_3$ symmetry operation, i.e., a rotation in the TMDC plane by an angle $\varphi=\pm2\pi/3$, which brings about a chiral selection rule $m=1$ for the $K$-point and $m=-1$ for the $K^\prime$-point, i.e., the chirality of the SPs matches that of the valley. For eccentric positions, which are not too far from $\mathbf r=0$, it is not exactly the case but still there is a preference for the chirally-matched SPs.

We assume that the pumping is performed with the circularly polarized radiation, and one of the valleys, say the $K$ valley, is predominantly populated. Consequently, the first mode that can go into generation is the $m=1$ SP. To find the necessary condition that the corresponding threshold can be achieved,  we will follow Ref.\ \onlinecite{Stockman_JOPT_2010_Spaser_Nanoamplifier} and set $n_\mathbfcal K=1$. Then from Eqs.\ (\ref{12}) and (\ref{14}) we obtain this condition as

\begin{equation}
\nu_\mathbf K \int_S \left|I_m^{(\pm)}(\mathbf r)\right|^2\mathrm d^2\mathbf r\ge 1~,
\label{Threshold}
\end{equation}
where $\pm$ is the chirality of the pumped TMDC valley, and the coupling amplitude is
\begin{equation}
I_m^{(\pm)}(\mathbf r)=\frac{A_m d_\mathbfcal K}{\hbar \sqrt{\gamma_\mathrm{sp}\Gamma_{12}}} \nabla\phi_m(\mathbf r)\mathbf e_\pm~.
\end{equation}
Note that $I_m^{(\pm)\ast}=I_{-m}^{(\mp)}$.

This coupling amplitude is illustrated in Fig.\ \ref{SP_Coupling} for $I_m^{(+)}$. As we see, for the chiral-matched SP with $m=1$, the coupling amplitude is approximately constant, $I_m^{(+)}\approx 1$ within the geometric footprint of the silver spheroid, which is seen in panel (a) as an almost uniform orange disk with radius $a=12$ nm. In a sharp contrast, the chiral-mismatched mode with $m=-1$ [panel (b)] has virtually no coupling to the TMDC transitions in the $K$ valley. Thus, for the gain medium whose radius $R_\mathrm g$ is within the footprint of the metal spheroid, i.e., for $R_\mathrm g\le a$, only one mode with chirality $m=1$ will be generated. However, for a larger gain medium, $R_\mathrm g\ge 13$ nm, there is a circle of strong coupling of the mismatched mode, which can potentially go into the generation.

 \subsection{Kinetics of Continuous-Wave Spasing}
 \label{Kinetics}
 
Below in this Article, we provide numerical examples of the spaser kinetics. For certainty, we assume that  the $K$-valley is selectively pumped, which can be done with the right-hand circularly polarized pump radiation. (As protected by the $\mathcal T$-symmetry, exactly the same results are valid for the left-handed pump and $K^\prime$-valley.) Thus, we set $g_K=g$ and $g_{K^\prime}=0$.

 A continuous wave (CW) solution can be obtained by solving Eqs.\ \eqref{12}-\eqref{14} where the time derivatives in the left-hand sides are set to zero.  The calculated dependences of the generated coherent SP population, $N_m=\left|a_m\right|^2$ where $m=\pm1$, on the pumping rate, $g$, for various TMDC's are shown in  Fig.\ \ref{fig:CWS.jpg}(a). As we can see, there is a single spasing threshold for each of the TMDCs. Significantly above the threshold, for $g_\mathcal K>30~\mathrm{ps^{-1}}$, the number of SPs, $N_m$, grows linearly with pumping rate $g$. This is a common general property of all spasers: it stems from the fact that the feedback in the spasers is very strong due to the extremely small modal volume. Therefore, the stimulated emission dominates the electronic transitions between the spasing levels, which is the prerequisite of the linear line $N_m(g)$. The slope of this straight line (the so-called slope efficiency) is specific for every given TMDC. For all these spasers, the threshold condition of Eq.\ (\ref{Threshold}) for the generation of the matched mode is satisfied. We have verified that the mismatched mode ($m=-1$) does not have a finite threshold, i.e., it is not generated at any pumping rate. The reason is that the matched mode ($m=1$) above its threshold clamps the inversion at a constant level \cite{Stockman_JOPT_2010_Spaser_Nanoamplifier} preventing its increase with the pumping and, thus, precluding the generation of the mismatched mode. In this case, the single chiral mode generation enjoys a strong topological protection. 


\begin{figure}
	\begin{center}
		\includegraphics[width=.95\columnwidth]{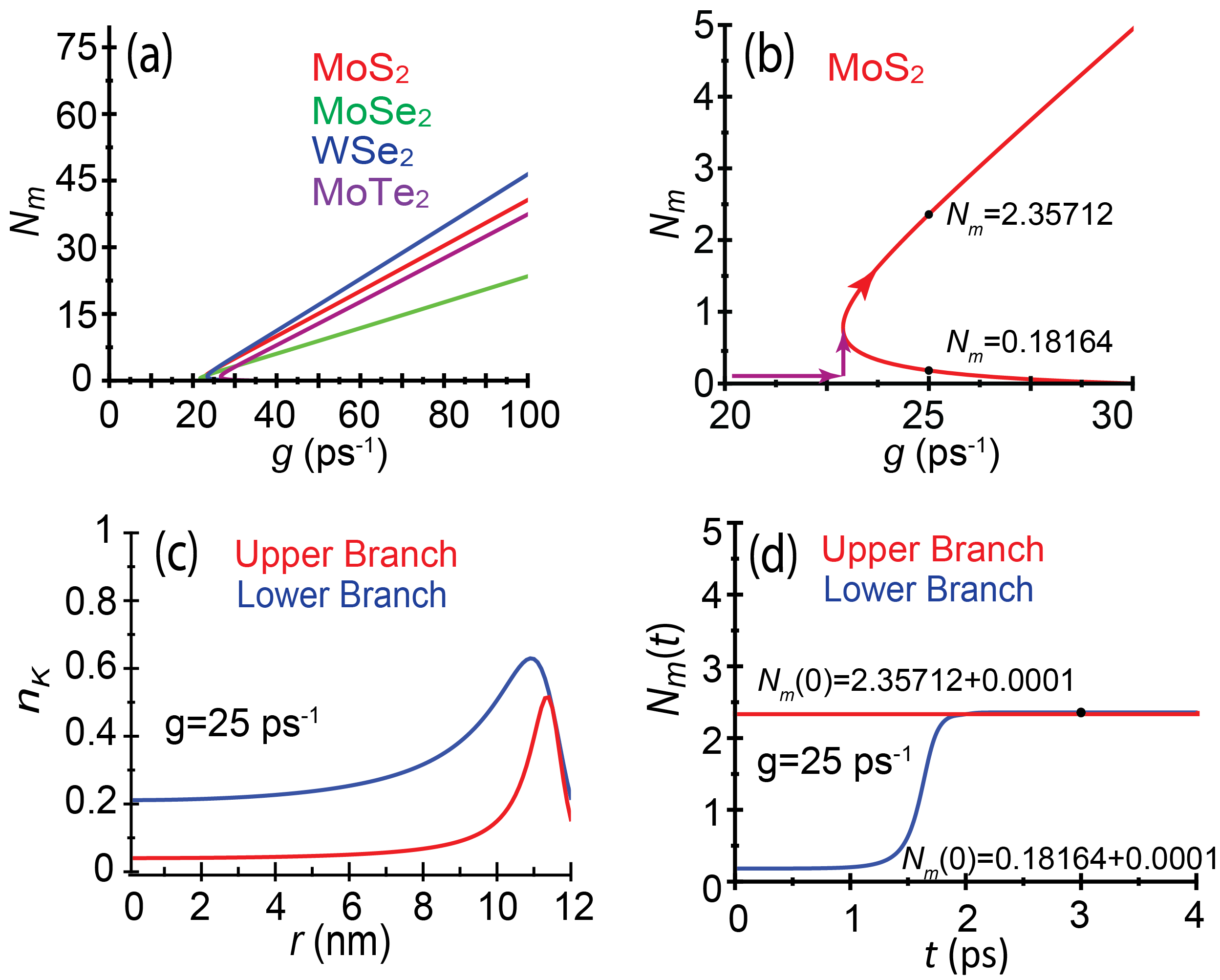}
		\caption{Spaser kinetics. (a) Dependence of the number of SP quanta in the spasing mode on the pumping rate for gain medium of the matched radius, $R_\mathrm g=a=12$ nm. Only the chirality-matched SP with $m=1$ are generated. (b) Magnified near-threshold portion of panel (a) for MoS$_2$. The number of the SPs, $N_m$, is indicated for the points shown on the graphs for the two branches.  (c) Radial distribution of the inversion, $n_k$ for each of the two branches. (d) Test of stability of the two SP branches. The kinetics of the SP population, $N_m$, after the number of the SPs in each branch is increased by $\Delta N_m=0.0001$.   }
		\label{fig:CWS.jpg}
	\end{center}
\end{figure}

At the threshold, the spasing curves experience a bifurcation behavior. This is clearly seen in the magnified plot in Fig.\ \ref{fig:CWS.jpg}(b): there is the threshold as the bifurcation point and two branches of the spasing curve above it. As we see from  Fig.\ \ref{fig:CWS.jpg}(c), these two branches differ by the stationary values of population inversion $n_\mathbfcal K$: for the upper branch it is significantly lower than for the lower branch. To answer a question whether these two branches are stable, we slightly perturb the accurate numerical solutions at $g=25~\mathrm{ps^{-1}}$ by changing the number of SPs by $\Delta N_m=0.0001$. The density matrix solution for the dynamics of the SP population induced by such a perturbation is shown in Fig.\ \ref{fig:CWS.jpg}(d). As we see, the upper branch is absolutely stable but the lower branch is unstable, and it evolves in time towards the upper branch within less than half a picosecond. As a result of this bifurcation instability, the system actually evolves with the increase of pumping along a path indicated by arrows in  Fig.\ \ref{fig:CWS.jpg}(b): Below the threshold, the population of the coherent SPs $N_m=0$; it jumps to the apex of the curve at the bifurcation point and then follows the upper branch. One can state that the spatial inhomogeneity of the field and the inversion cause the spasing transition to become the first order. This is in contrast to the previous homogeneous case of Ref.\ \onlinecite{Stockman_JOPT_2010_Spaser_Nanoamplifier} where this transition was continuous, i.e., of the second order.

\begin{figure}
	\begin{center}
		\includegraphics[width=.95\columnwidth]{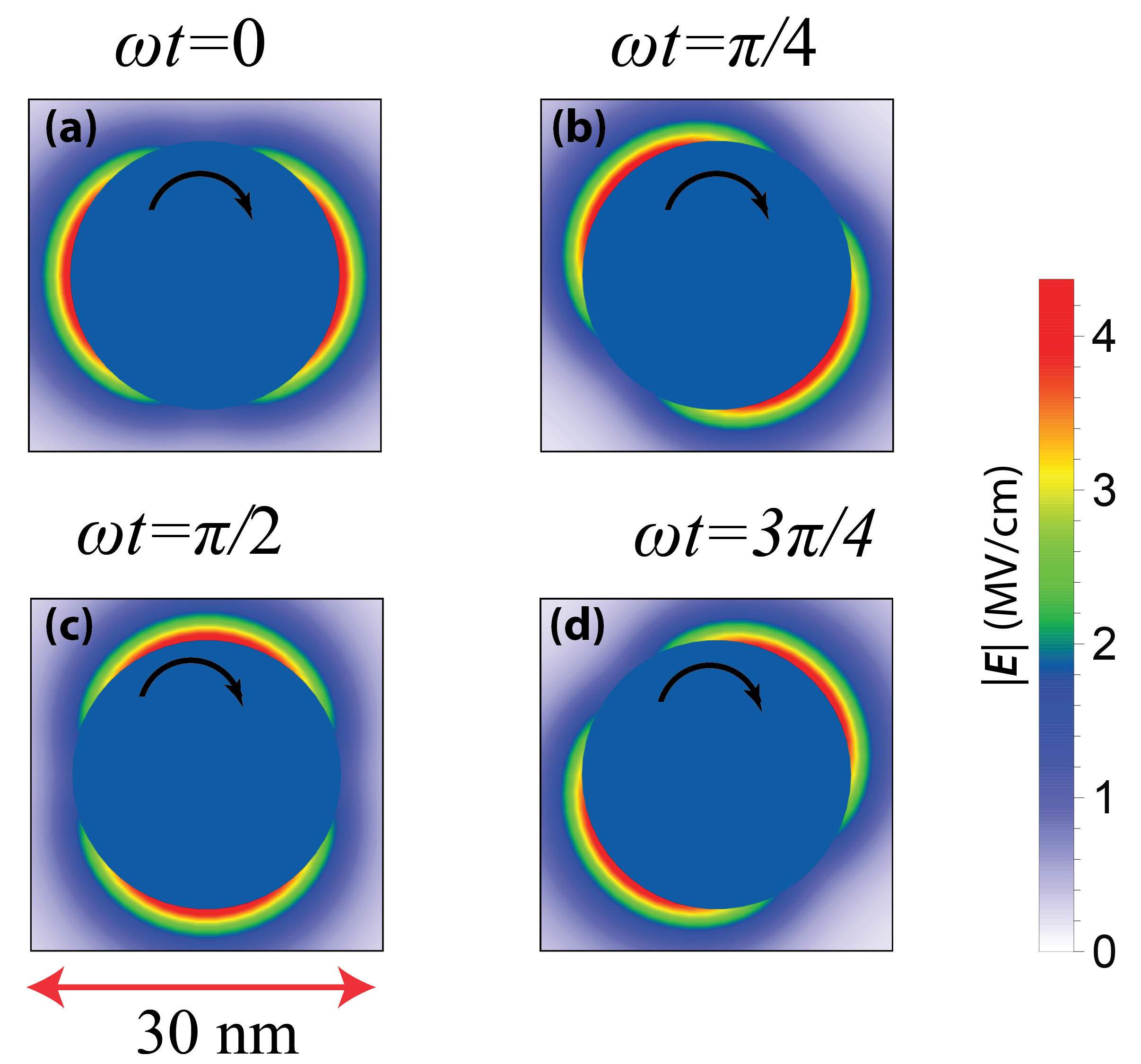}
		\caption{Temporal dynamics of the local electric field, $\left | \mathbf E\right|$, in topological spaser generating in the $m=1$ mode. The curved arrow indicates the rotation direction of the field (clockwise). The magnitude of the field is calculated for a single SP per mode, $N_m=1$; it is color-coded by the bar to the right. The phase of the spaser oscillation is indicated at the top of the corresponding panels.}
		\label{fig:bist4}
	\end{center}
\end{figure}

The chiral optical fields generated by the topological spaser are not stationary -- they evolve in time rotating clockwise for $m=1$, as illustrated in Fig.\ \ref{fig:bist4}, and counterclockwise for $m=-1$. The magnitude of the field is large even for one SP per mode, $\left | \mathbf E\right|\sim 10^7~\mathrm{V/\AA}$, which is a general property of the nanospasers related to the nanoscopic size of the mode. Note that with increase of the SP population, the field increases as $\left | \mathbf E\right|\propto \sqrt{N_m}$.

\begin{figure}[h]
	\begin{center}
		\includegraphics[width=.95\columnwidth]{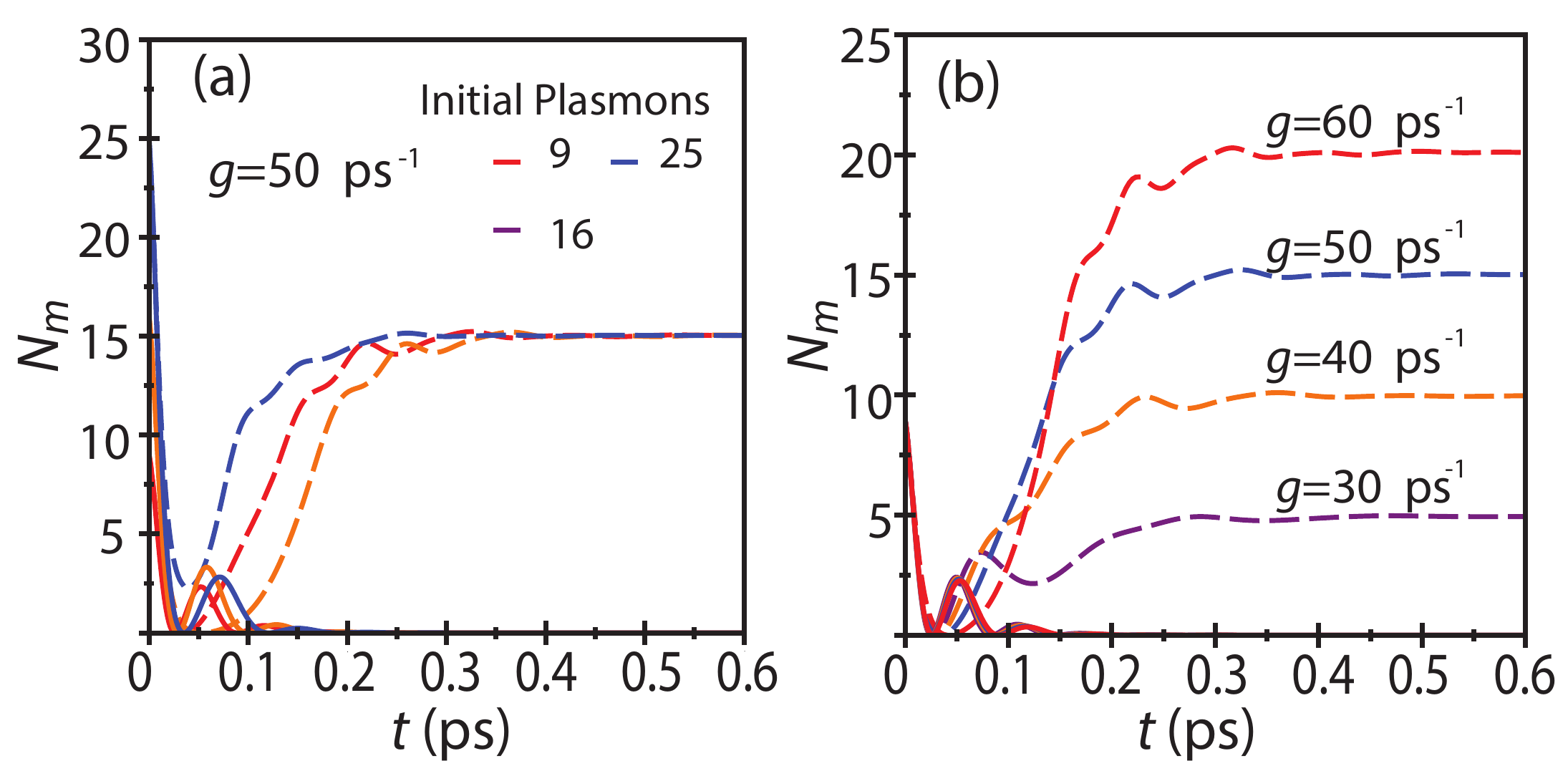}
		\caption{Number of SPs $N_m$ as a function of time $t$ for a spaser with MoS$_2$ as a gain material. The pumping is performed by a radiation whose electric field rotates clockwise in the plane of system ($m=1$). The solid lines denote the  chiral  SPs with $m=-1$, and the dashed lines denote the SPs with $m=1$. The pumping rates are indicated in the panels. (a) Dependence of SP number $N_m$ on time $t$ after the beginning of the pumping for different initial SP populations (color coded as indicated) for pumping rate $g=50~\mathrm{ps^{-1}}$. (b) Dependence of SP number $N_m$ on time $t$  for different pumping rates $g$ (color coded). The initial SP number is $N_m=10$.}
		\label{diffg}
	\end{center}
\end{figure}

\subsection{Stability and Topological Protection of Spaser Modes}
\label{Stability}

In Fig.\ \ref{diffg}(a), we test the stability and topological protection of the spasing mode. Panel (a) displays the dynamics of the SP population of the topological spaser, $N_m(t)$, for different initial numbers of SPs, $N_m(0)$, and for their different chiralities, $m=\pm1$. As these data show, the left-rotating SPs ($m=-1$) are not amplified irrespectively of their initial numbers: the corresponding curves evolve with decaying relaxation oscillations tending to $N_{-1}=0$. In contrast, the $m=1$ SPs exhibit a stable amplification: their number increases to a level that is defined by the pumping rate, $g$, and does not depend on the initial populations. The $m=1$ chirality SP-amplification stability with respect to the injection of the $m=-1$ quanta, which these data demonstrate is due to the topological protection: matching the phase windings of the SP mode and the electronic states in the pumped $K$-valley.

As a complementary test, we show in Fig.\ \ref{diffg}(b) the temporal dynamics of the SP population for equal initial number of SPs but different pumping rates. The dynamics in this case is again stable with the mismatched $m=-1$ SPs decaying to zero, and the matched $m=1$ being amplified to the stable levels that linearly increase with the pumping rate.

\subsection{
Far-Field Radiation of Spaser}
\label{Radiation}

The spaser is a subwavelength device design to generate intense, coherent nanolocalized fields. Generation of far-field radiation is not its primary purpose. However, the proposed spaser, as most existing nanospasers, generates in a dipolar mode that will emit in the far field. This emission, in absolute terms, can be quite intense for a nanosource. In particular, the spaser emission was used to detect cancer cells in the blood flow model \cite{Galanzha_Nat_Comm_Spaser_biological_probe_2017}; it was, actually, many orders of magnitude brighter than from any other label for biomedical detection.

To describe the spaser emission, we note that the radiating dipole uniformly rotates with the angular velocity of $\omega_\mathrm{sp}$. The emitted radiation will be right-hand circularly polarized for the pumping at the $K$ point and left-hand circularly polarized for the $K^\prime$ pumping. Note that the corresponding two radiating modes are completely uncoupled. This is equivalent to having two independent chiral spasers in one.

To find the intensity, $I$, of the emitted radiation, we need to calculated the radiating dipole. To do so, we will follow Ref.\ \onlinecite{Stockman_Opt_Expres_2011_Nanoplasmonics_Review}.   We take into account that the modal field, $\mathbf E_m=\nabla\phi_m$, inside the metal spheroid is constant. Then from Eq.\ (8) of the SM, we can find  
\begin{equation}
E_m^2=\frac{s_\mathrm{sp}}{V_\mathrm m}~,
\label{E_m}
\end{equation}
where $V_\mathrm m$ is the spheroid's volume. The physical field squared inside the metal is found from Eqs.\ (\ref{Fm}) and (\ref{Asp}),
\begin{equation}
F_m^2=\frac{4\pi\hbar s_\mathrm{sp}^2 N_m}{\epsilon_\mathrm d s^\prime_\mathrm{sp} V_m}~.
\label{FFm}
\end{equation}
From this, we find the radiating dipole squared as 
\begin{equation}
\left|d_\mathrm{0p}\right|^2=\frac{\hbar}{4\pi}\left(\mathrm{Re}\frac{\partial \epsilon_\mathrm m(\omega_\mathrm{sp})}{\partial \omega_\mathrm{sp}}\right)^{-1} \mathrm{Re}\left[\epsilon_\mathrm m(\omega_\mathrm{sp})-\epsilon_\mathrm d\right]^2 V_\mathrm m N_m~.
\end{equation}
The dipole radiation rate (photons per second) can be found from a standard dipole-radiation formula \cite{Landau_Lifshitz_Field_Theory} as 
\begin{eqnarray}
I&=&\frac{4}{9} \left(\frac{\omega}{c_0}\right)^3 \left(\epsilon_\mathrm d\right)^{1/2} \mathrm{Re}\left[\epsilon_\mathrm m(\omega_\mathrm{sp})-\epsilon_\mathrm d\right]^2 \times
\nonumber\\
&&\left(\mathrm{Re}\frac{\partial \epsilon_\mathrm m(\omega_\mathrm{sp})}{\partial \omega_\mathrm{sp}}\right)^{-1} a^2 c N_m~,
\label{I}
\end{eqnarray}
where $c_0$ is speed of light in vacuum.

For our example of MoS$_2$, substituting parameters that we used everywhere in our  calculations [see Sec.\ \ref{Params}], we obtain 
\begin{equation}
I=2.1\times 10^{12}N_m~\mathrm{s^{-1}};~~~P=\hbar\omega_\mathrm{sp} I=0.55N_m~\mathrm{\mu W}~,
\end{equation}
where $P$ is the power of the emission. From these numbers, we conclude that the emission is bright for a nano-emitter and easily detectable. This is in line with the observation of the emission from single spasers of the comparable size in Ref.\ \onlinecite{Galanzha_Nat_Comm_Spaser_biological_probe_2017}.

\section{Concluding Discussion}
\label{Conclusion}

In this Article we introduce a topological nanospaser that consists of a plasmonic metal spheroid as the SP resonator and a nanoflake of a semiconductor TMDC as a gain medium. This spaser has two mutually $\mathcal T$-reversed dipole modes with identical frequencies but opposite chiralities (topological charges $m=\pm1$). Only the mode whose chirality matches that of the active (pumped) valley, i.e., $m=1$ for the $K$-valley and $m=-1$ for the $K^\prime$ valley, can be generated while the conjugated mode does not go into generation at any pumping level. The topological spaser is stable with respect to even large perturbations: the SP with mismatched topological charge injected into the system even in large numbers decay exponentially within a $\sim100$ fs time. This implies a strong topological protection. 

This protection is not trivial because the exact valley selection rule matching its chirality to the of the SPs is strictly valid only on the symmetry axis of the metal spheroid (in the center of the TMDC gain medium flake). Off-axis, there is a coupling of the gain to the chirally-mismatched SPs. However, the strong topological protection appears due to the fact that the spaser is a highly-nonlinear, threshold phenomenon. In fact, it is the nonlinear saturation of the gain and the concurrent clamping of the inversion that cause the strong mode competition. The topologically-matched mode ($m=1$ for the $K$-valley and $m=-1$ for the $K^\prime$ valley) reaches the threshold first and saturates the gain, thus, preventing the mismatched mode from the generation under any pumping or any perturbations.

The proposed topological spaser is promising for the use in nanooptics and nanospectroscopy where strong rotating nano-localized fields are required. It may be especially useful in applications to biomolecules and biologicals objects, which are typically chiral. It may also be used as a nano-source of a circularly polarized radiation in the far field, in particular, as a biomedical multi-functional agent similar to the use of spherical spasers \cite{Galanzha_Nat_Comm_Spaser_biological_probe_2017}.

\begin{acknowledgments}
Major funding was provided by Grant No. DE-FG02-11ER46789 from the Materials Sciences and Engineering Division of the Office of the Basic Energy Sciences, Office of Science, U.S. Department of Energy. Numerical simulations have been performed using support by
Grant No. DE-FG02-01ER15213 from the Chemical Sciences, Biosciences and Geosciences Division, Office of Basic Energy Sciences, Office of Science, US Department of Energy. The work of V.A. was supported by NSF EFRI NewLAW Grant EFMA-17 41691. Support for J.W. came from a MURI Grant No. N00014-17-1-2588 from the Office of Naval Research (ONR).
\end{acknowledgments}

\section{Appendix}
\subsection{Modes of a metallic oblate spheroid}
For an oblate spheroid, geometry is defined by the spheroidal coordinates, $\xi$, $\eta$ and $\varphi$ related to the Cartesian coordinates, $x$, $y$ and $z$ as \cite{Willatzen_Voon_2011_Book_Boundary_Problems}:
\begin{align}
x&=f\sqrt{\xi^2+1}\sqrt{1-\eta^2}~ \cos(\varphi)\label{1},\\
y&=f \sqrt{\xi^2+1}\sqrt{1-\eta^2}~\sin(\varphi)\label{2},\\
z&=f \xi \eta\label{3},
\end{align}
 with  $ 0 \leq \xi < \infty$, ~ $ -1 \leq \eta \leq 1$, ~ $ 0 \leq \varphi < 2 \pi$ and $f > 0$.
The corresponding equation in the Cartesian coordinate system describes a spheroidal shape  with the semi-principal axes $a$ and $c$,
\begin{equation} 
\dfrac{x^2+y^2}{a^2}+\dfrac{z^2}{c^2}=1~.
\end{equation}
The eccentricity, $\varepsilon$, can be written as $\varepsilon=\sqrt{1-\frac{c^2}{a^2}}$ and $f= \varepsilon a$.

The surface plasmon eigenmodes of the spheroid are described by the quasistatic equation \cite{Stockman:2001_PRL_Localization}
\begin{align}
\nabla\left[\theta(\mathbf{r})\nabla \phi_{m} \right]= s_{\mathrm{sp}} \nabla^2 \phi_{m}. 
\end{align}
Here $\theta(\mathbf{r})$ is the characteristic function that is equal to 1 inside the metal and 0 elsewhere. 
For the spheroid, the multipole quantum number is $l=1$, and $m$ is the angular momentum projection. Eigenmodes are given by
\begin{align}
\phi_m =C_\mathrm{N}  P_{1}^{m}(\eta) e^{im\phi}
\begin{cases}
 \frac{P_1^{m}(i\xi)}{P_1^{m}(i\xi_{0})},  &  0<\xi<\xi_{0},\\
 \frac{Q_1^{m}(i\xi)}{Q_1^{m}(i\xi_{0})}, &  \xi_{0}<\xi,
\end{cases}\label{boundary}
\end{align}
where $ P_l^{m}(x)$ and $ Q_l^{m}(x)$ are the Legendre functions of the first and second kind, respectively, and 
$\xi_0 = \frac{\sqrt{1-\varepsilon^2}}{\varepsilon}$.
 The constant $C_{N}$  is determined by the normalization condition of the eigenmodes,
\begin{align}
\int_{\mathrm{All~Space}}  |\nabla \phi(\mathbf{r})_m|^2 d^3 \mathbf{r}=1.
\end{align}

Due to the axial symmetry, the corresponding eigenvalues, $s_\mathrm{sp}$, do not depend on $m$; they are given by the equation \cite{Bergman_Stockman:2003_PRL_spaser, Stockman_JOPT_2010_Spaser_Nanoamplifier}
\begin{align}
s_\mathrm{sp} = \frac{\displaystyle{ \int_{\mathrm{All~Space}}  \theta(\mathbf{r}) |\nabla \phi_m(\mathbf{r})|^2 d^3 \mathbf{r}}}{\displaystyle{ \int_{\mathrm{All~Space}}  |\nabla \phi_m,(\mathbf{r})|^2 d^3 \mathbf{r}}}.
\end{align}
From this, we derive an analytical form of the eigenvalue 
\begin{align}
	s_\mathrm{sp} =\left. \frac	{\frac{d P_1^m(x)}{dx}} {\frac{d P_1^m(x)}{dx}  - \frac{P_1^m(x)}{Q_1^m(x)} \frac{d Q_1^m(x)}{dx}	}	\right|_{x=i\xi_0}~.
	\label{s_m}
\end{align}
	To find the SP frerquency, $\omega_{\mathrm{sp}}$, and the SP relaxation rate, $\gamma_{\mathrm{sp}}$, we use relations  \cite{Bergman_Stockman:2003_PRL_spaser, Stockman_JOPT_2010_Spaser_Nanoamplifier}
\begin{align}
&s_{\mathrm{sp}} = \mathrm{Re}[s(\omega_{\mathrm{sp}})], \\
&r_{\mathrm{sp}} = \frac{\mathrm{Im}[s(\omega_{\mathrm{sp}})]}{s_{{\mathrm{sp}}}^{\prime}},~~~~
s_{{\mathrm{sp}}}^{\prime}\equiv \frac{d\mathrm{Re}[s(\omega)]}{d\omega}\Big|_{\omega=\omega_{\mathrm{sp}}},
\end{align}
where the Bergman spectral parameter is defined as 
\begin{align}
s(\omega) = \frac{\epsilon_{d}}{\epsilon_{d}-\epsilon_{m}(\omega)}.
\end{align}
$\epsilon_d$ is the dielectric constant of the environment, and $\epsilon_m(\omega)$ is the dielectric function of  the metal. In our computations, we use silver with the dielectric function from the Ref.  \onlinecite{Johnson:1972_Silver}.

\subsection{TMDC Parameters}

\begin{figure}
	\begin{center}
		\includegraphics[width=.95\columnwidth]{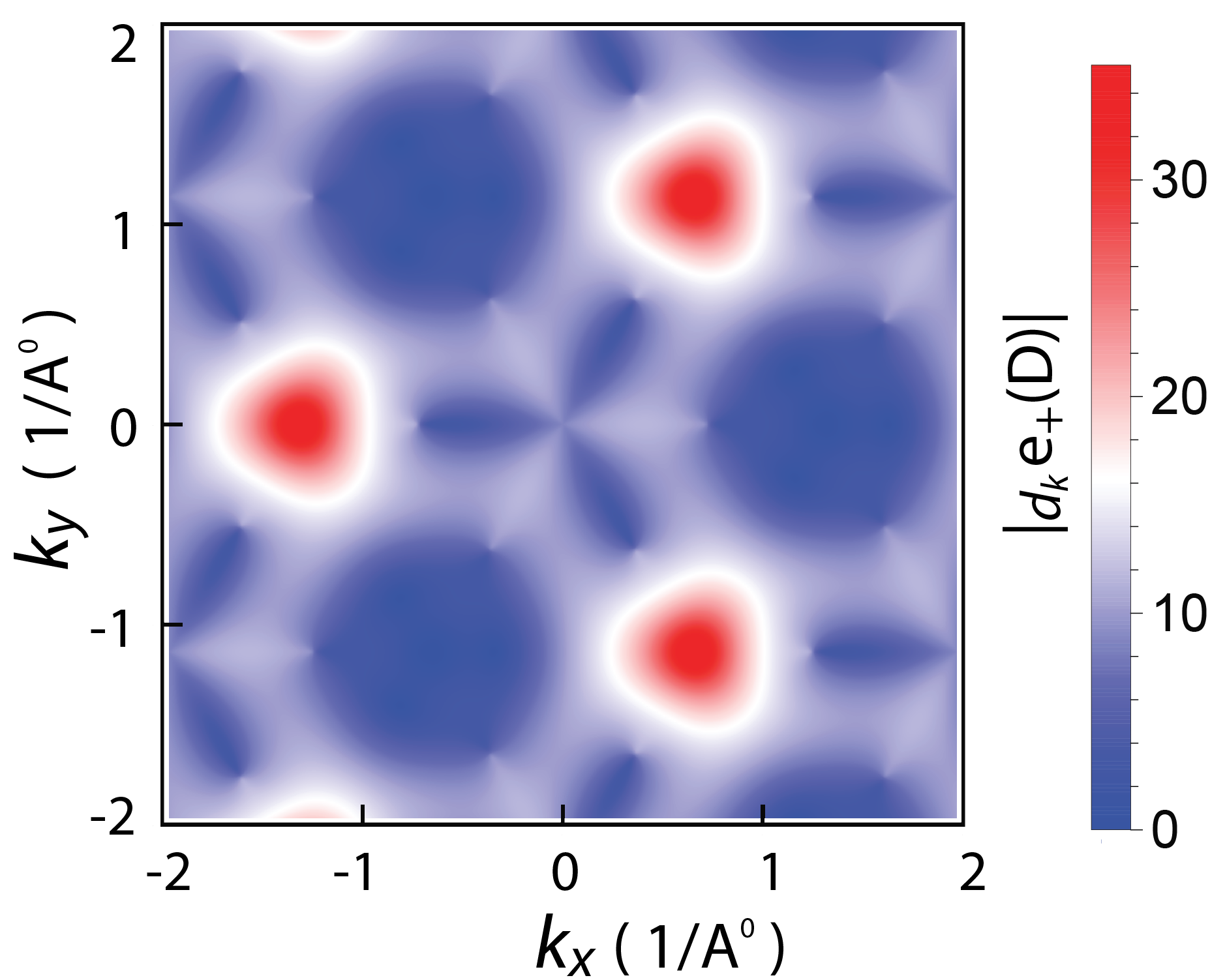}
		\caption{Absolute value of the left-rotating chiral dipole component, $\mathbf d_-=\mathbf e_+\mathbf d$, in MoS$_2$}
		\label{fig:bist4.eps}
	\end{center}
\end{figure}

The eigenfrequency, $\omega_\mathrm{sp}$, depends on the aspect ratio, $c/a$. We set $a=12$ nm and vary the value of $c$ to have the surface plasmon frequency match the band gap, $ \omega_\mathrm{sp} = \Delta_\mathrm{g}$. Table \ref{Table} shows the calculated  values of $c$ to match the band gaps of the specific TMDCs.

The transition dipole matrix element of the gain mediaum (TMDC) were calculated using a three-band tight binding model \cite{Liu_et_al_PRB_2014_Three_Band_Model}. The calculated values of the transition dipoles  along with the band gap are indicated in Table \ref{Table}. The transition dipoles at the $K$ and $K^\prime$ points are purely chiral: they are proportional to $\mathbf e_\pm=2^{-1/2}\left(\mathbf e_x\pm i\mathbf e_y\right)$, where $\mathbf e_x$ and $\mathbf e_y$ are the Cartesian unit vectors. A plot of absolute value of the chiral dipole, $\left|\mathbf d_\pm\right|$, where $\mathbf d_\pm=
\mathbf e_\pm^\ast \mathbf d$,  is shown in Fig.\ref{fig:bist4.eps}. It clearly points to the chirality of transition dipoles ($\mathbf d_+$  at the $K$ point and $\mathbf d_-$  at the $K$ point) in the TMDCs. 

\begin{table}
\begin{center}
	\begin{tabular}{ | c|>{\centering}p{2.2cm}| >{\centering}p{1.8cm}  |>{\centering}p{1.8cm} | c| l |}
\hline
		\multirow{2}{*} {TMDC} & 
		{Semi-principal axis } &
		 \multicolumn{2}{c|}{Dipole elements (D)} &
		 {Band gap} \\ 
\cline{3-4}
		                               &  $c$ (nm) &  $\mathbf d_\mathbf K$  &  $\mathbf d_{\mathbf K^{\prime}}$ &(eV) \\ \hline
		$\mathrm{MoS}_2$   & $1.20$     &  $17.68 \mathbf{e}_{+}$  &  $17.68\mathbf{e}_{-}$                    &  1.66 \\ \hline
		$\mathrm{MoSe}_2$  & $1.45$    &  $19.23 \mathbf{e}_{+}$  &  $19.23\mathbf{e}_{-}$                     &1.79 \\ \hline
		$\mathrm{WSe}_2$   &  $0.85$   &  $18.38 \mathbf{e}_{+}$  &  $18.38\mathbf{e}_{-}$                      &1.43 \\ \hline
		$\mathrm{MoTe}_2$  &  $1$       &  $20.08 \mathbf{e}_{+}$  &  $20.08\mathbf{e}_{-}$                      & 1.53 \\ \hline
	\end{tabular}
\caption{Parameters employed in the calculations: Semi-principal axes of the spheroids, and the dipole matrix elements and band gaps of the TMDCs.}\label{Table}
\end{center}
\end{table}

\end{document}